%% file: main.tex
\documentclass[conference]{IEEEtran}
\IEEEoverridecommandlockouts
\usepackage{cite}
\usepackage{amsmath,amssymb,amsfonts}
\usepackage{algorithmic}
\usepackage{algorithm}
\newtheorem{definition}{Definition} 
\usepackage{graphicx}
\usepackage{textcomp}
\usepackage{xcolor}
\def\BibTeX{{\rm B\kern-.05em{\sc i\kern-.025em b}\kern-.08em
    T\kern-.1667em\lower.7ex\hbox{E}\kern-.125emX}}
\begin{document}

\title{Efficient Memory Management for GPU-based Deep Learning Systems}
\author{\IEEEauthorblockN{Junzhe Zhang, Sai Ho Yeung, Yao Shu, Bingsheng He, Wei Wang}
\IEEEauthorblockA{\textit{School of Computing, National University of Singapore} \\
\{junzhe, yeungsh, shuyao95, hebs, wangwei\}@comp.nus.edu.sg}}

\maketitle
\begin{abstract}

GPU (graphics processing unit) has been used for many data-intensive applications. Among them, deep learning systems are one of the most important consumer systems for GPU nowadays. As deep learning applications impose deeper and larger
models in order to achieve higher accuracy, memory management becomes an important research topic for deep learning systems, given that GPU has limited memory size. Many approaches have been proposed towards this issue, e.g., model compression and memory swapping. However, they either degrade the model accuracy or require a lot of manual intervention.
In this paper, we propose two orthogonal approaches to reduce the memory cost from the system perspective. Our approaches are transparent to the models, and thus do not affect the model accuracy. They are achieved by exploiting the iterative nature of the training algorithm of deep learning to derive the lifetime and read/write order of all variables. With the lifetime semantics, we are able to implement a memory pool with minimal fragments. However, the optimization problem is NP-complete. We propose a heuristic algorithm that reduces up to 13.3\% of memory compared with Nvidia's default memory pool with equal time complexity. With the read/write semantics, the variables that are not in use can be swapped out from GPU to CPU to reduce the memory footprint. We propose multiple swapping strategies to automatically decide which variable to swap and when to swap out (in), which reduces the memory cost by up to 34.2\% without communication overhead.
\end{abstract}


\section{Introduction}
\label{intro}
GPU has boosted the performance of many data-intensive applications, including database management~\cite{he2011high,He:2008:RJG:1376616.1376670,7930016,Wang:2014:CAQ:2732967.2732976,He:2009:RQC:1620585.1620588,atung18}, graph processing~\cite{Sha:2017:ADG:3151113.3151122}, and machine learning  tasks~\cite{8440051,krizhevsky2012imagenet}. Among them, deep learning, also known as deep neural networks (DNNs), is one of the most successful and popular applications of GPU. The training algorithm of DNNs involves many large matrix production operations. By accelerating the matrix operations via thousands of processing units in parallel, GPU enables us to train complex DNN models efficiently, speeding up the training for one order of magnitude \cite{dogaru2017, scanzio2010}. Numerous studies have shown that larger and deeper DNNs can significantly increase the model accuracy \cite{bahrampour2011comp, ngiam2011on} for computer vision and natural language processing applications. However, GPU has limited memory while DNNs are memory hungry. For instance, the AlexNet \cite{krizhevsky2012imagenet} was trained on two GPUs (each with 3 Giga Bytes) in parallel to overcome the memory limitation, while the VGG network is much larger and has to be trained on a 4-GPU system \cite{simonyan2015}. This limitation has been a bottleneck to explore deep and wide DNNs to capture complex regularities of the big data~\cite{bahrampour2011comp}. In fact, there are many system challenges~\cite{wang2016database,re2015machine,li2018ease} for deep learning. In this paper, we focus on memory optimization from the system perspective.

Various techniques in reducing GPU memory footprint have been proposed (see Section~\ref{related_work} for the details), including (i) buffering and paging; (ii) model compression, (iii) memory sharing, (iv) trading computation for memory, and (v) memory swapping. However, those approaches have their respective drawbacks or limitations. For instance, the general buffering and paging strategy work at the coarse memory granularity, which is not optimal in memory saving for deep learning whose variables' sizes vary significantly. The model compression approaches tend to decrease the accuracy or introduce quantization error~\cite{vanhoucke2011improv}. Existing swapping approaches require a lot of human intervention.

Towards these issues, we propose two automatic approaches for efficient and effective GPU memory optimization that can be easily adopted into existing deep learning systems and are transparent to the end-users. They are also orthogonal to multiple-GPU systems that partition the model or data onto GPUs to reduce the memory footprint of each GPU. Our approaches do not alter the model structure or training algorithm; hence, there is no effect on the accuracy and convergence. They do not require computational graph semantics or knowledge of specific DNN models, but exploit the iterative nature of the deep learning training algorithms to derive the lifetime and read/write order of all variables for memory optimization. In particular, the first approach implements a smart GPU memory pool that optimizes the memory allocation based on the lifetime of all variables. For example, variables without overlapping in lifetime can be allocated into the same memory space, i.e. memory sharing. However, finding the optimal allocation scheme is an NP-complete problem. In this paper, we propose a heuristic solution to solve it. 

The second approach automatically swaps variables not in use from GPU to CPU memory and swaps them back before the next access. We observe that the back-propagation algorithm for training DNNs has a special pattern that variables from the bottom layers are only accessed at the beginning and end of each iteration. Consequently, the swapping approach has the potential to significantly save the GPU memory. However, it incurs communication and synchronization overhead. The swapping schedule, that decides which variable to swap and when to swap, has to be designed carefully to hide the overhead. We propose multiple scheduling strategies to trade off the overhead and the memory reduction.

The contributions of this paper include:
\begin{itemize}
\item We propose two approaches to reduce the GPU memory cost of deep learning training, including a memory pool and an automatic swapping mechanism. 

\item We implement the two approaches under a unified abstraction, which collects the lifetime and read/write orders of the variables, and then runs the memory pool and memory swapping.

\item We conduct experiments to evaluate the performance of our approaches. The results confirm the superiority of our approaches against baselines. In particular, our memory pool reduces up to 13.3\% of memory compared with Nvidia's default memory pool called CnMem with equal time-complexity. Our memory swapping approach reduces memory cost by up to 34.2\% without incurring any communication overhead.
\end{itemize}
The rest of this paper is organized as follows. We review the related work in Section~\ref{related_work}. Two optimization algorithms for memory pool management and memory swapping are introduced in Section \ref{memory pool management} and Section~\ref{auto-swap} respectively, followed by their implementations in Section \ref{implementation}. The performance of the proposed methods is evaluated in Section~\ref{evaluation}. We conclude this paper in Section~\ref{conclusion and future work}.

\section{Related Work and Background}
\label{related_work}
Due to the limited GPU memory size, effective memory management is a must for handling large data sets on GPUs. In this section, we review related work on general and GPU memory management, then we give some background information of the training algorithm of DNNs to introduce the iterative nature. There are two terms used frequently in this section, namely, \textbf{memory footprint} and \textbf{memory load}:

\begin{definition}
  \textbf{Memory footprint} or \textbf{memory consumption} is the actual GPU memory consumed by the program. They are used interchangeably in this paper. 
\end{definition}
\begin{definition}
    \textbf{Memory load} at a point of time is the summation of sizes of all the variables which are currently residing in the GPU memory. Memory load does not consider memory allocation, such as the memory imposed by buffers, fragmentation, etc.
\end{definition}

\subsection{Memory Management}
Memory management is an important component of computer systems, including database systems. Various techniques have been proposed such as paging and cache optimization~\cite{zhang2015memory,palkar2018evaluating}.
Memory pool is one effective memory optimization technique. It pre-allocates a continuous chunk of memory and takes over the memory management from the operating system. There are numerous variations of memory pool implementations~\cite{alger2000c++}, in order to obtain small time complexity and \textbf{competitive ratio}~\cite{bender2017cost}. 
CnMem\cite{cnmempool} is a GPU memory pool developed by Nvidia. However, it does not have optimization for the training of deep learning. In this paper, we propose to optimize the memory allocation within the GPU memory pool by exploiting variables' lifetime and size information to achieve a better competitive ratio with low time complexity.

\begin{definition}
    \textbf{Competitive ratio} is the ratio of memory footprint to \textbf{peak memory load}.
\end{definition}
\begin{definition}
    \textbf{Peak memory load} is the maximum of the memory load in a training iteration. The logical time where the peak memory load occurs is defined as the \textbf{peak time}.
\end{definition}

\subsection{GPU Memory Management}
There have been proposals on system supports for paging on the GPU~\cite{Ausavarungnirun:2017:MGM:3123939.3123975, 7446077}. However, those approaches usually require modifications on hardware or drivers. In data processing, many existing studies (e.g.,~\cite{He:2009:RQC:1620585.1620588,Sha:2017:ADG:3151113.3151122}) partition the data into chunks and ensure that the processing of each data chunk can fit into the GPU memory. This approach is effective only if the data accesses are regular and partitionable into chunks. The variables' sizes of DNNs vary a lot, which requires the optimization algorithm to work at a more fine-grained granularity, e.g., variable, than page or buffer. Wang et al.~\cite{Wang:2014:CAQ:2732967.2732976} proposed a cost-driven replacement policy on the GPU memory, which combines the effects of data size, eviction latency, and data locality. However, those studies do not take advantage of the iterative nature of deep learning training, which we demonstrate to significantly improve the effectiveness of memory management for deep learning systems.

\subsection{GPU Memory Management for Deep Learning }
\subsubsection{Model Compression}
Reducing the precision of parameters or the complexity of the network structure reduces the memory load in deep learning. Half precision (16FP), single bit, and mixed precision numbers have been applied in DNNs~\cite{ courbariaux2014low, le2013spfp, mcdonnell2018training}. 
Typically, using low precision numbers introduces quantization error~\cite{vanhoucke2011improv}. Model pruning that removes the layers or connections in the DNNs~\cite{hassibi1993second, han2016effi} reduces the complexity and memory load; however, it also decreases the training accuracy. It should be noted that our proposed solutions are orthogonal to the model compression approaches stated above. 

\subsubsection{Memory Sharing}
Two types of memory sharing have been proposed for deep learning frameworks, which are in-place operation and buffer reuse. In-place operation is to store the output at the physical address of the input. For instance, $y$ can be stored at the memory block of $a$ directly when computing $y=sigmoid(a)$. There are only a few in-place operations in DNNs. Buffer reuse is to share the memory between variables whose lifetime does not overlap, as implemented in MXNet-memonger~\cite{chen2016training}. It can be conveniently implemented into deep learning frameworks of declarative paradigm (e.g. TensorFlow, MXNet), where the entire computation graph is constructed before computing, providing the topology and data dependencies for smarter memory allocation \cite{chen2015mxnet, chen2016training}. It should be noted that the SmartPool approach proposed in this paper provides superior (more fine-grained) memory sharing, which will be elaborated in Section~\ref{memory pool management}.

\subsubsection{Trading Computation for Memory}
Some intermediate variables, such as feature maps, are freed during forward-propagation and recomputed during backward-propagation to compute gradients. Relevant works are seen in MXNet-memonger \cite{chen2016training}, SuperNeurons \cite{wang2018superneurons}, and the DenseNets implementation via PyTorch \cite{pleiss2017memory}, all of which have the cost-awareness idea to selectively drop the feature maps which are easy to recompute. However, recomputing variable requires high-level semantics, i.e. computation graph and hence cannot be done at the memory management level.

\subsubsection{Memory Swapping}
Memory load can be reduced by swapping variables to CPU memory when they are not in use, and swapping them back to GPU memory right before their next access. Ideally, the communication in both directions should be hidden under computation (via separate streams) to minimize the communication overhead \cite{he2011high}.
In GeePS \cite{cui2016geeps}, the decision to swap which layer or which tensor is made by the end user. It requires the end user to have a good understanding of the model, including the memory and time consumption of each layer. SuperNeurons \cite{wang2018superneurons} restricts to swap only convolution layers. Big tensors of other layers are not considered for swapping. It also requires either the computational graph of the DNN or the end-user's intervention. 
It is desired to provide a fully automatic approach that is transparent to end-users and is able to swap memory with small communication overhead.

\subsection{The Iterative Training Algorithm}
\label{iterative nature}
\begin{figure}
  \includegraphics[scale=0.32]{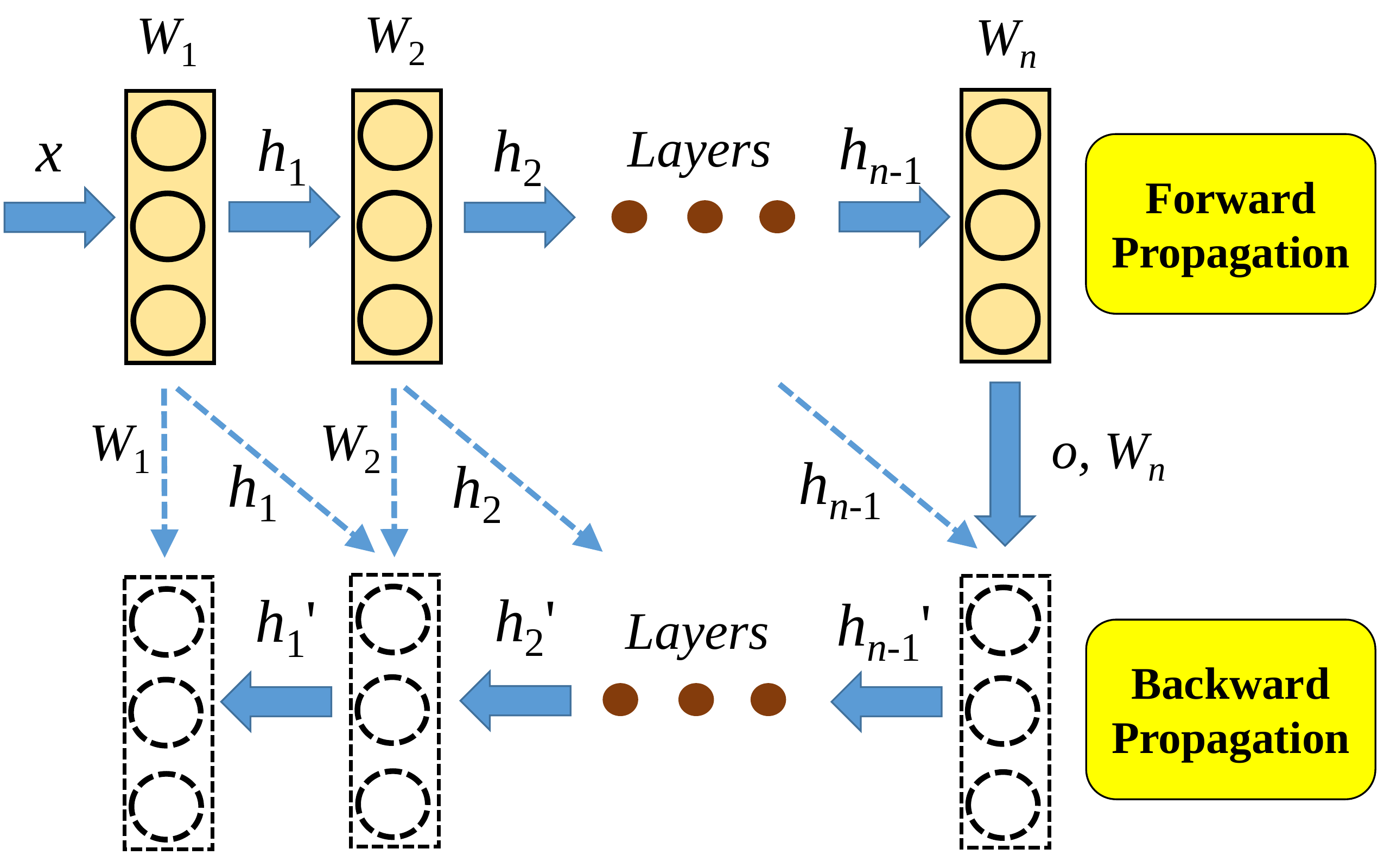}
  \caption{Forward-propagation and backward-propagation.}
  \label{fig:ff_bp}
\end{figure}

\begin{figure}
  \includegraphics[scale=0.53]{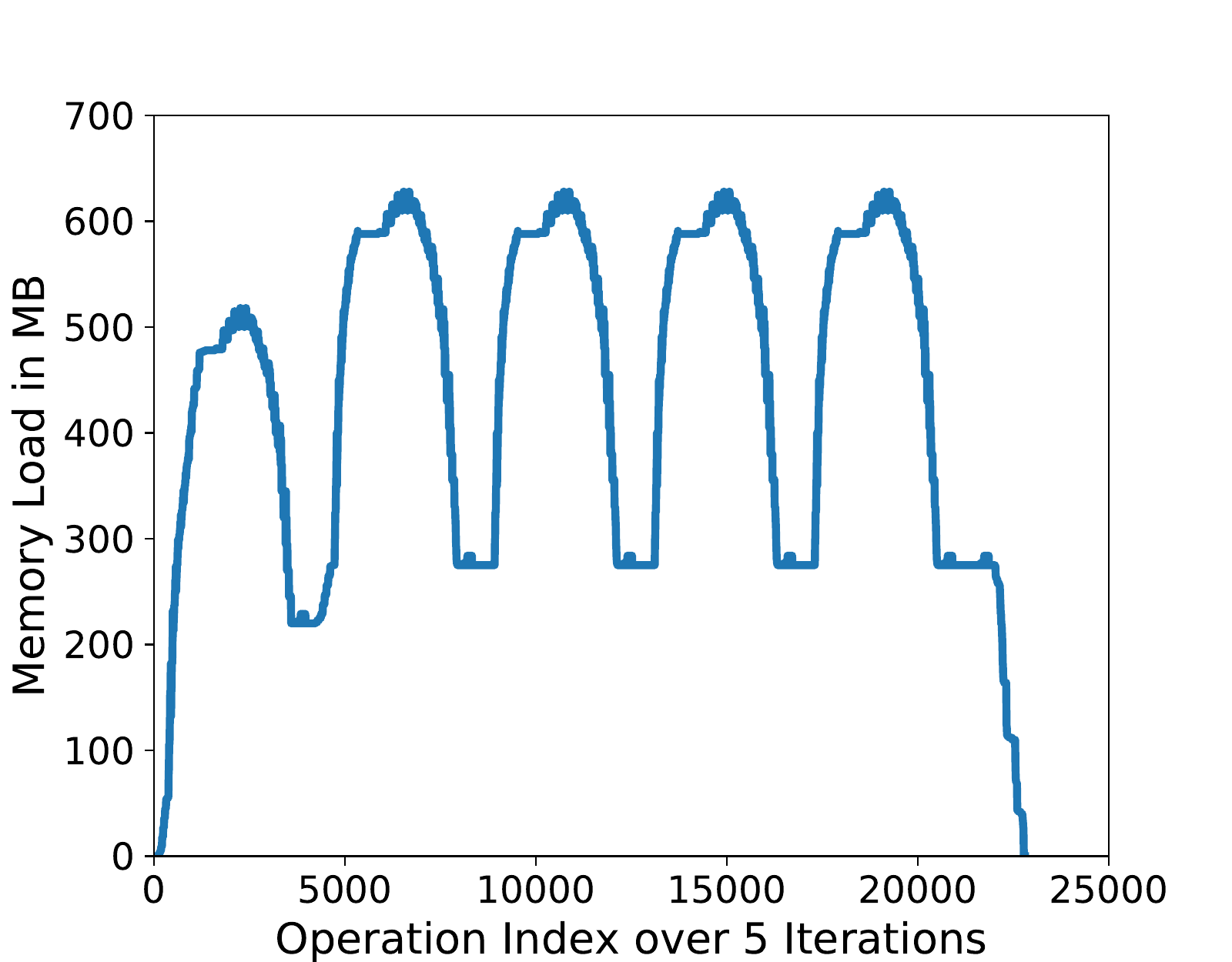}
  \caption{Memory load for VGG16 in 5 iterations.}
  \label{fig:load_iter}
\end{figure}

Many machine learning algorithms such as DNNs are iterative~\cite{lecun2015deep}.  The stochastic gradient descent (SGD) algorithm for training DNNs typically needs thousands of iterations (or even more) to converge. In each iteration, the network undergoes forward-propagation and backward-propagation as illustrated in Fig. \ref{fig:ff_bp}, during which up to tens of thousands of variables of different size are created, read, updated and deleted. The network computes feature maps, i.e., $h_1, h_2,\cdots, h_{n-1}$, during forward-propagation, obtains the loss at the end of forward-propagation, and then computes gradients to update weights during backward-propagation. Feature maps computed in forward-propagation are normally retained in GPU memory until they are used to compute weights gradients. Hence, the overall memory usage increases during forward-propagation, peaks at the end of forward-propagation, and decreases during backward-propagation. The memory load profile over a 5-iteration training process for the VGG network \cite{simonyan2015} is shown in Fig. \ref{fig:load_iter}, with each operation index referring to one malloc / free / read / write operation. We observe that lifetime, sizes, and read/write sequences of the variables are rather stable across all the middle iterations (from the second to the second last) as shown in Fig.~\ref{fig:load_iter}, for most DNNs except stochastic neural networks \cite{turchetti2004}. This iterative nature enables us to collect the lifetime and read/write semantics of variables at the beginning iterations, and optimize memory management for the later iterations (in Sections~\ref{memory pool management} and \ref{auto-swap}).


\section{Memory Pool: SmartPool}
\label{memory pool management}
In this section, we exploit the iterative nature of the training algorithm of DNNs to optimize the allocation of variables in a memory pool. We call the proposed memory pool SmartPool.

\subsection{Dynamic Storage Allocation (DSA)}
\begin{figure}
    \includegraphics[scale=0.55]{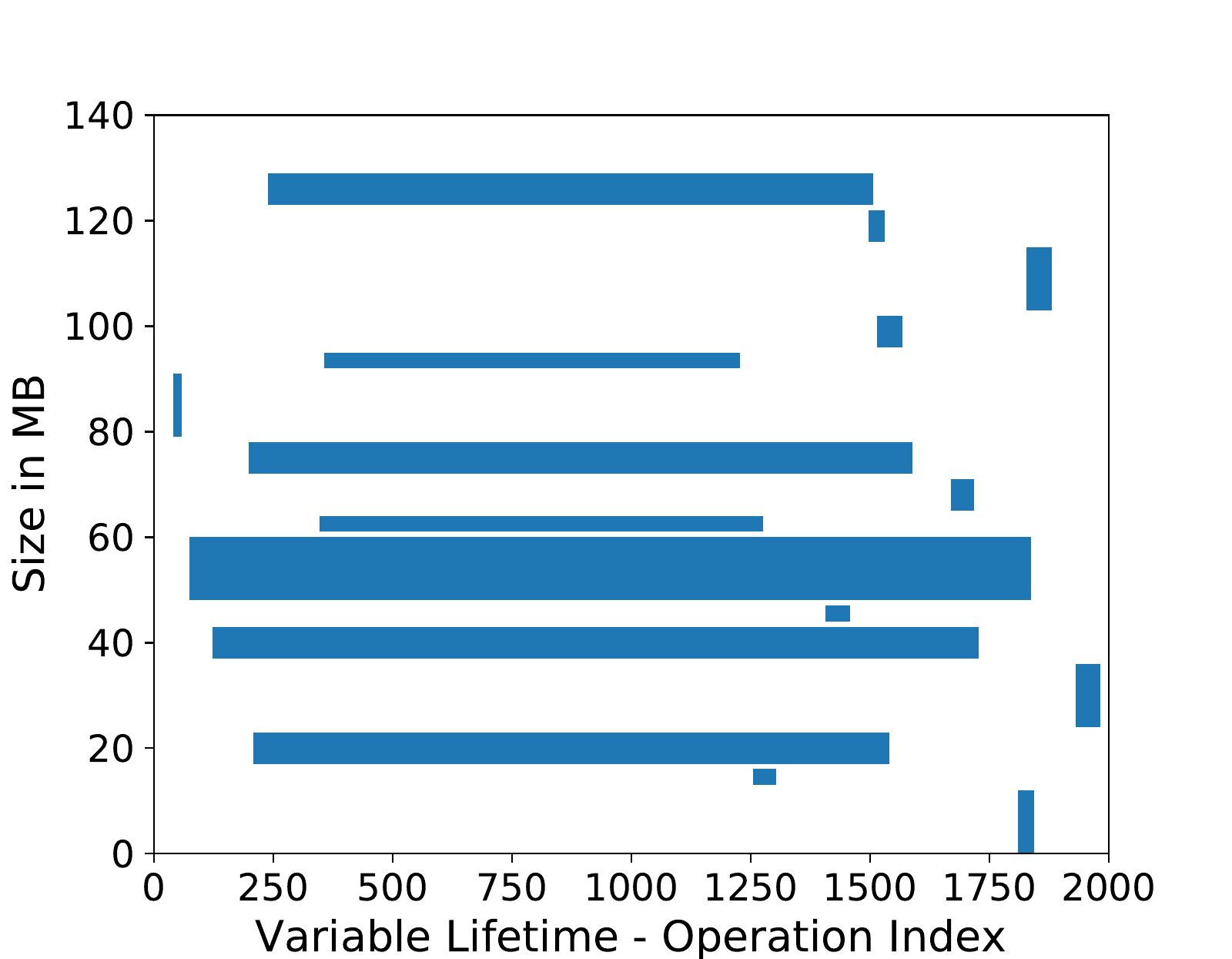}
  \caption{Illustration of the lifetime and size of variables.}
  \label{fig:prob_state}
\end{figure}

Over one training iteration, there are tens of thousands of variables allocated and freed at the fixed logical time, i.e., the operation indices. This can be illustrated in a 2D view shown in Fig. \ref{fig:prob_state}, where the $x$-axis represents the logical time in one iteration, and the $y$-axis represents the memory location in Mega Bytes (MB). Each rectangle refers to a variable, where the width and the height refer to the lifetime and size of the variable respectively. The lifetime information tells at which operation indices a variable is allocated and then deallocated. 
Optimization of the memory footprint can be done by moving the rectangles vertically such that some rectangles without overlapping in lifetime can share the same memory, i.e., overlapping vertically. 
The problem of allocating and deallocating memory blocks has been well studied as Dynamic Storage Allocation (DSA)~\cite{bender2017cost}. DSA is an NP-complete problem which was first shown by Stockmeyer \cite{garey1976some, kierstead1991polynomial}. Depending on the semantics, the problem can be categorized as: (i) online DSA, where items must be allocated once they arrive without the information of the items yet to arrive; and (ii) offline DSA, where the lifetime and size of all variables are known before allocating the first object. Offline DSA obviously provides more semantics and normally is able to achieve better performance~\cite{lu2016lifetime}.
The iterative nature of deep learning enables us to collect the lifetime and size of all variables in one iteration, and hence the memory footprint optimization problem is an offline DSA problem.

\subsection{Weighted Graph Representation of Offline DSA}
\label{graph coloring algorithm}

We follow the graph coloring
methods to solve the offline DSA problem in the deep learning context, which guarantees desirable time complexity, competitive ratio, and scalability. A method named Weighted Interval Color (WIC) \cite{kierstead1991polynomial} was first developed, which provides a good representation of the offline DSA problem by adopting the weight parameter $W$ into the graph, denoted by $G(V, E, W)$.
 
In this representation, vertices $V$ denotes the memory blocks storing the variables. Edges $E$ denotes the pairwise relationship between vertices, with $E_{ij}=1$ representing that there is overlapping in lifetime of two memory blocks (i.e. vertices) $i$ and $j$, or $0$ otherwise. The weights $W$ here are on the vertices rather than edges, denoting the sizes of the memory blocks.

The weighted interval coloring procedure assigns a range of discrete color values to each node (i.e., variable) with the length equal to its weight instead of one color value, such that the color range is never shared with any of its neighbors. The color range assigned is, in fact, the range of memory on the GPU RAM where a corresponding memory block resides.

In interval graph coloring, clique number, denoted as $\omega(G)$, refers to the largest clique size in a graph, which is equivalent to the smallest number of colors required for coloring. In the context of weighted interval coloring, $\omega(G)$ becomes:

\begin{equation}
    \omega(G) = max_{clique\ j=0}^{m}\left(\sum_{vertex\ i=0\ in\ clique\ j}^{n} w_i\right)
\end{equation}

$\omega(G)$ here is actually the peak memory load defined in Section \ref{related_work}. The chromatic number $\chi(G)$ of WIC is the number of colors assigned to the graph by this algorithm.
It is actually equivalent to the memory footprint of this allocation, and is bounded between 1 and $\alpha$ times $\omega(G)$, with $\alpha$ being the competitive ratio in the DSA problem.
\begin{equation}
    \omega(G) \leq \chi(G) \leq \alpha * \omega(G)
\end{equation}

\begin{figure}
    \includegraphics[scale=0.52]{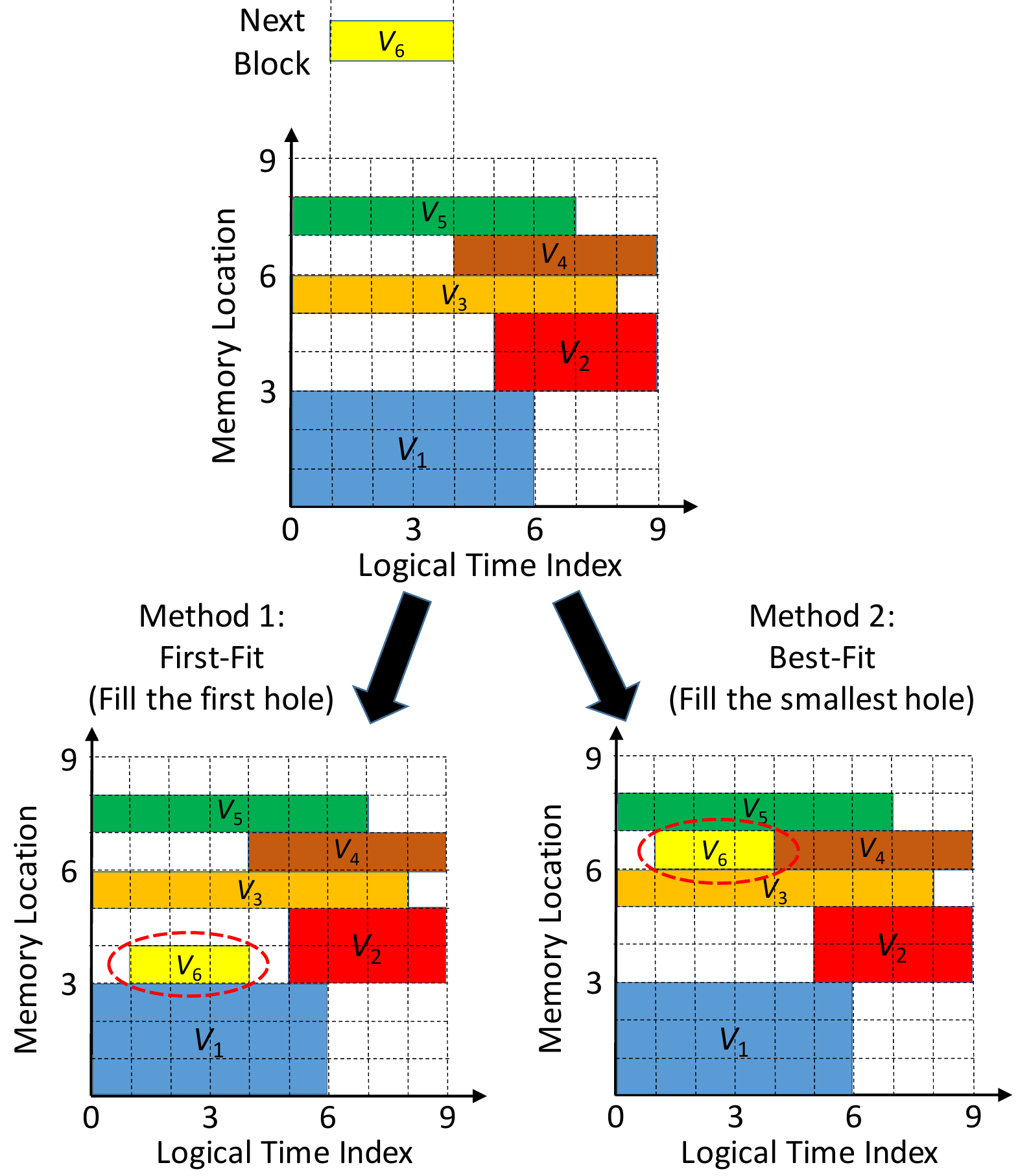}
  \caption{Illustration of the difference between first-fit and best-fit.}
  \label{fig:first_best_fit}
\end{figure}

\subsection{SmartPool Algorithm}
\label{allocation}

In this section, we present our heuristic algorithm that implements WIC~\cite{kierstead1991polynomial}. Our algorithm drops the idea of rounding each weight to a power of two, which saves memory space by up to half. 

Given a list of variables with their pairwise time-overlapping relationship, we first sort these variables in the descending order of size. Starting from the largest variable, we assign a memory block for each variable such that the memory block does not overlap (in lifetime) with any of its neighbors that had been allocated to other variables. In particular, the algorithm tries to fit the variable into one of the free holes between preoccupied memory blocks. There are two methods of fitting blocks into the memory pool, which are first-fit and best-fit. Both are simple yet effective \cite{pemmaraju2004approximating}. 
To demonstrate the idea, the difference between first fit and best fit is illustrated in Fig.~\ref{fig:first_best_fit}. In the figure, when a variable is to be allocated by the memory pool, the first-fit method will allocate it to the first hole that the variable can fit in, while the best-fit method will allocate it to the smallest hole where the variable can fit in. Furthermore, in the case that the memory pool does not have such a big hole for the variable, the memory pool will be extended and the new memory will be used to for the variable. As a result, the memory footprint increases as allocation progresses.

In our implementation, best-fit is the default option. Note should be taken that the algorithm only needs to run once when constructing the memory pool; the real allocation is done by looking up a hash table, which will be elaborated in Section \ref{eval-smartpool}. Hence, the overhead introduced by best-fit is negligible compared to the entire training process. 

In previous work, one variable shares memory with only another variable (one-to-one). In SmartPool, as can be seen from Fig. \ref{fig:first_best_fit}, memory can be shared among those blocks(variables) without overlap in lifetime and this sharing is not restricted to one-to-one sharing. SmartPool allows more superior sharing such that one large variable can share memory with several small ones (whether consecutive or not), and vice versa.

\section{Memory Swapping: AutoSwap}
\label{auto-swap}

Each variable during its lifetime might be accessed multiple times for reading and writing. For example, in Fig.~\ref{fig:ff_bp}, the variables from the bottom layers are accessed at the beginning and the end of each training iteration. These variables could be swapped to CPU memory after one access to reduce the memory load, and then prefetched before the next access. However, swapping incurs communication and synchronization overhead, which would slow down the training speed. Even though we can perform the swapping using 2 separate cudaStreams (one for swap-out and the other for swap-in) in parallel to the computation cudaStream, the communication and synchronization overhead may not be completely hidden under computation. In this section, we propose AutoSwap to schedule the swapping in order to minimize the overhead and maximize the reduction of memory load.

It applies simple filters to get a set of candidate variables (Section~\ref{sec:candidate}) for swapping. These candidate variables are ranked according to priority scores (Section~\ref{PS}); Given the GPU memory limit, AutoSwap schedules (Section~\ref{selection}) the swapping by selecting the variables with higher rankings.

\subsection{Candidate Swapping Variables}\label{sec:candidate}
We set up two simple criteria to obtain a list of candidate variables for swapping. First, we filter out tiny variables that are smaller than a threshold, e.g.
, 1 MB. Because small variables like the bias vectors in the DNNs, have little effect in memory load reduction if they are swapped out. Moreover, we have to bear the cost of the latency in data transmission. Second, we only consider swapping the variables whose lifetime are across the peak time. There are two advantages: (i) The memory load is most tight near the peak time and hence most eagerly to be reduced; (ii) Intermediate variables whose lifetime spans across only a few layers during forward-propagation or backward-propagation are excluded, since they have to be swapped back to GPU in a short while. In other words, swapping these variables contributes to memory reduction for a very short period. Moreover, it occupies the I/O (i.e., PCIe) bandwidth.

\begin{figure}
  \includegraphics[scale=0.38]{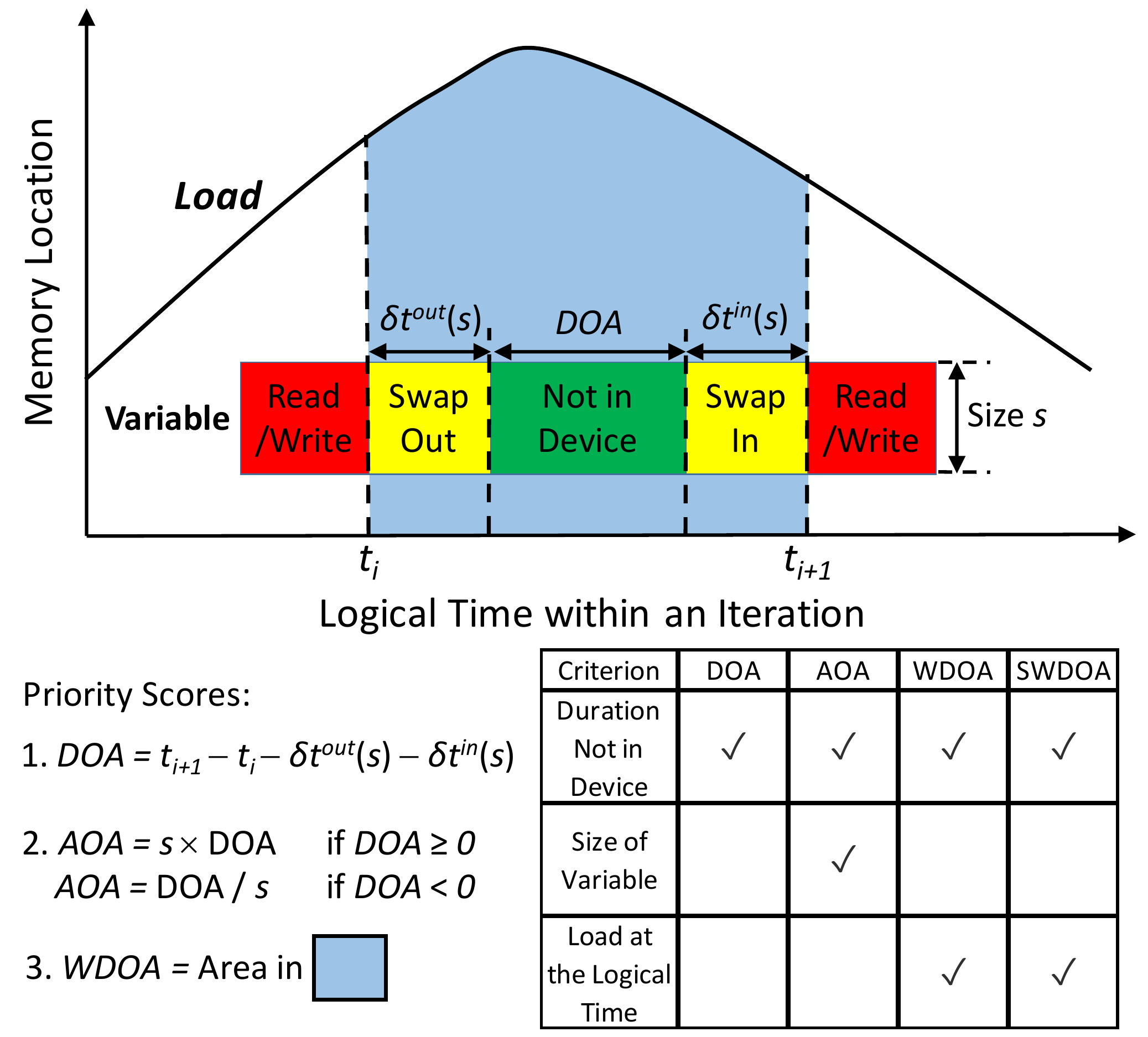}
  \caption{Illustration of the swapping priority scores for variables.}
  \label{fig:priority_score}
\end{figure}

\subsection{Priority Scores}
\label{PS}
According to the analysis in Section~\ref{sec:candidate}, to evaluate if a variable is worth swapping, we shall look at: (i). the size of the variable, (ii). the time span that it is able to be absent from the GPU memory (from swap-out completes till swap-in starts) and (iii). the position of the time span within a training iteration. To quantify the relative priority of the candidate variables, we come up with a set of Priority Scores (PS) where each of the PS takes one or two factors into account.  The four PS are discussed below, with illustration shown in Fig. \ref{fig:priority_score}.

\textbf{(i) Duration of Absence (DOA):} 
It is simply the duration between two accesses minus the time spent in swap-out and swap-in. A candidate variable with larger DOA is preferred for swapping as the variable resides outside of the GPU for a longer period within one iteration. However, DOA does not care about the variable size, or where the absence is located. 

\textbf{(ii) Area of Absence (AOA):} 
It is the product of DOA and size, which quantifies the amount of memory load reduction over a time period. It can be viewed as removing an AOA amount of area below the memory load curve. A candidate variable with larger AOA is preferred for swapping as it reduces a larger portion of memory and/or for a longer period. However, it does not consider the location of the absence.

For large candidate variable in the top layers of the neural networks, DOA can be negative because the swap-in and swap-out time ($\delta_t^{out}+\delta_t^{in}$), i.e., data transfer time, may be larger than the period between two consecutive accesses of the same variable without swapping ($t_{i+1} - t_i$).
In this case, the definition of AOA is changed to the product of the inverse of variable size and DOA. As such, for the negative AOA, larger AOA represents a relatively higher contribution of memory load reduction. 

\begin{figure}
  \includegraphics[scale=0.5]{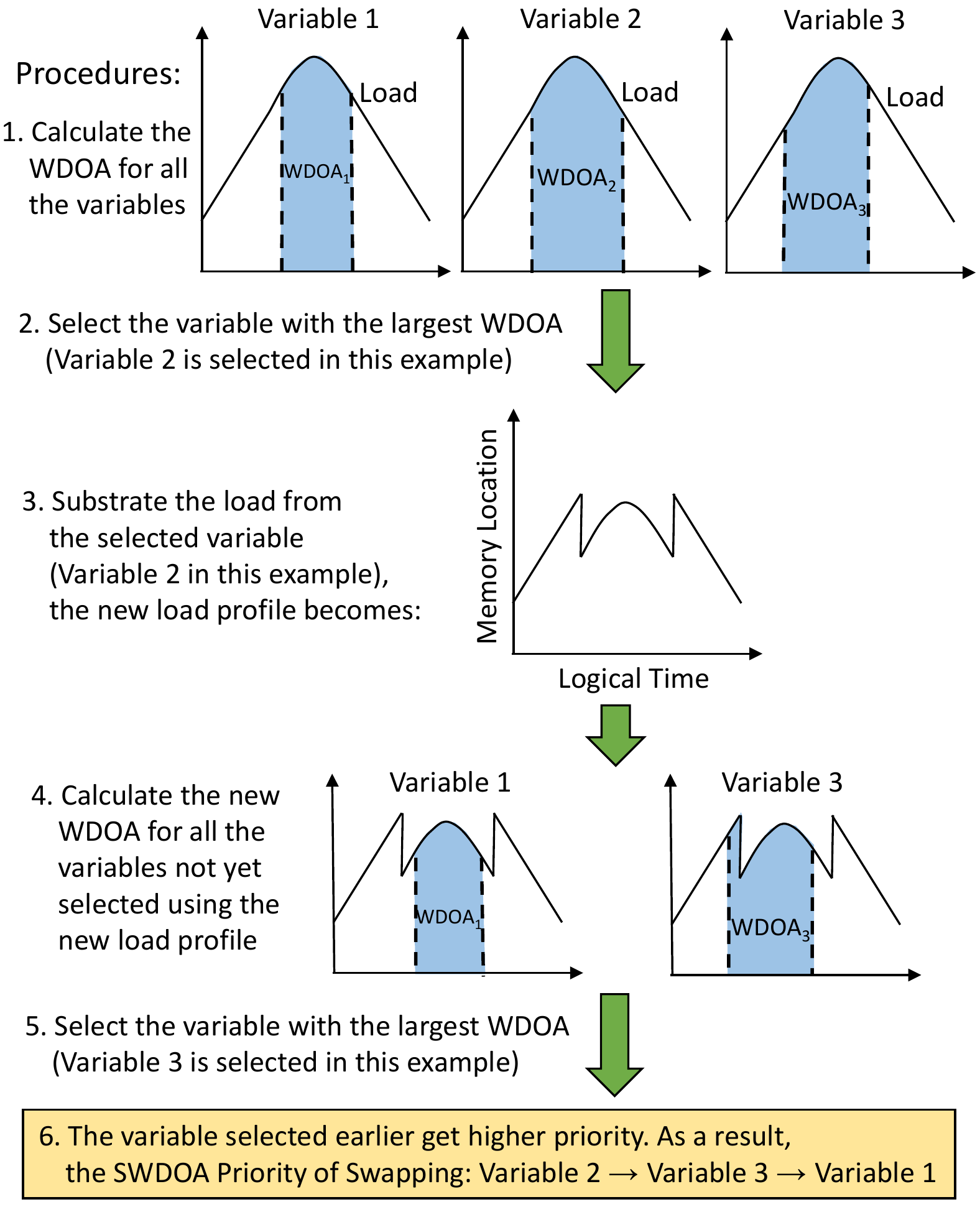}
  \caption{Illustration of the procedures to determine Submodular Weighted Duration of Absence (SWDOA).}
  \label{fig:sw_doa}
\end{figure}

\textbf{(iii) Weighted Duration of Absence (WDOA):}
WDOA of a candidate variable is the area under the original memory load curve between two consecutive accesses without swapping. It considers both DOA and the memory load during the period when the variable is swapped out, but not the variable size. A variable with larger WDOA is preferred as it indicates that this variable will be accessed in a longer time frame and the accesses are nearer the peak memory load. 

\begin{figure*}
  \includegraphics[scale=0.50]{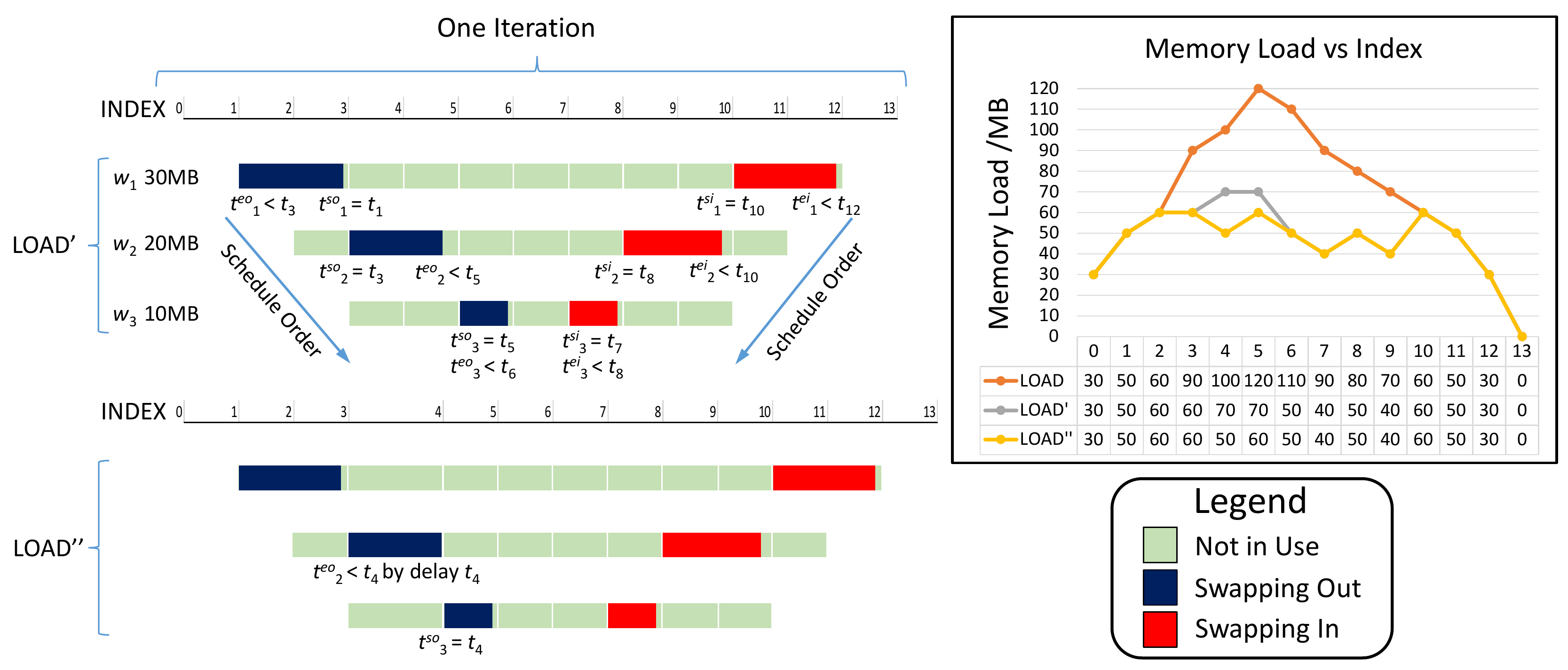}
  \caption{Illustration of scheduling.}
  \label{fig:illus}
\end{figure*}

\textbf{(iv) Submodular WDOA (SWDOA):}
WDOA for all candidate variables is computed based on the original memory load without swapping any variable. However, with the selection process progressing, WDOA for the remaining variables are inaccurate since the memory load changes due to that some variables are swapped.

SWDOA computes WDOA in a submodular way with updated memory load profile. It computes WDOA for all candidate variables under the original memory load, selects the variable with the highest WDOA, and updates the memory load. This process is repeated for all candidate variables using the updated memory load. An illustration of the SWDOA procedure to determine the swapping priorities of three variables are shown in Fig. \ref{fig:sw_doa}. SWDOA addresses the problem in which the peak time in the updated memory load profile shifts to somewhere else after a number of variables are selected. 
\subsection{Bayesian Optimization}

As discussed in Section \ref{PS}, none of the 4 PS are able to cover all the three factors. Moreover, as will be discussed in Section \ref{evaluation}, it is complex to find which factor is dominating in a particular scenario. Therefore, it is desirable to have a powerful PS that can combine the 4 existing PS. In this paper, we propose to exploit Bayesian Optimization to generate the aggregated priority score, denoted as BO. 

In particular, we generate the aggregated PS via a linear combination of the 4 basic scores, as shown in the equation below. The 4 priority scores are standardized before feeding into the formula. We use Bayesian Optimization to automatically tunes the weights $a$, $b$, $c$, and $d$, whose values range between [-1,1]. Gaussian Process prior is chosen \cite{snoek2012prac}, and the optimization objective is set to be the communication overhead. The weights are randomly initialized at the beginning. Then we run the swapping scheduling algorithm using the combined PS. The communication overhead is measured and fed into the Bayesian Optimization framework to update the weights. After updating about 30 times, the weights become stable, i.e., converge. Compared with the total training time that takes over thousands of iterations, the cost of the optimization procedure (about 30 iterations) is negligible.

\begin{equation}
    BO = a*AOA + b*DOA + c*WDOA + d*SWDOA
\end{equation}

\subsection{Swapping Variables Selection}
\label{selection}
To select the swapping variables, the PS or BO of all candidate variables are calculated. These variables are then sorted in descending order and inserted into a swapping list one by one according to the sorted order, while the memory load is updated (reduced) at the same time. When the peak memory load is no larger than the user-defined \textbf{memory load limit}, the selection stops.
\begin{definition}
    \textbf{Memory load limit}: In the AutoSwap approach, users define the value of memory load limit such that the peak memory load after swapping is no larger than this limit. 
\end{definition}

\subsection{Swapping Scheduling Algorithm}
\label{scheduling}
AutoSwap schedules to swap the selected variables out when they are not in use, and swap them in before the next access, i.e., prefetching. To maximize the PCI-e bandwidth, we schedule all the swap-out events in ascending order of their access time. A variable is swapped out immediately after the access completes. The swap-in events are scheduled according to the next access time of each variable. A variable starts swap-in in advance to make sure its second access is not delayed by the communication. The schedule is designed in the way such that a variable starts swap-out (-in) only after the previous swap-out (-in) event completes.

Communication and synchronization can be completely hidden under computation with relaxed memory load limit. However, under the condition of stringent memory load limit, memory load would exceed the limit due to that the variables cannot be swapped out quickly. If it occurs during forward-propagation, the next Malloc will be delayed until sufficient variables are swapped out and the memory is freed; if during backward-propagation, a swap-in can start only after freeing some variables. This is where communication and synchronization overhead occurs and is not able to be hidden. 

A simple example of the scheduling is illustrated in Fig. \ref{fig:illus}. In this example, the peak memory load is 120 MB and the memory load limit is 60 MB. 3 variables ($w_1,w_2,w_3$) are selected to meet the memory load limit. The scheduling is conducted based on above-mentioned strategy, where there are two types of time $t$ involved: the time with a superscript refers to the scheduled time of the $i$-th swapping variables, where the first superscript letters $s$ and $e$ represent the start and end of swap event respectively, and the second superscript letters $o$ and $i$ denote the swap out and swap in directions respectively; the time without a superscript refers to the time of $j$-th operation index. $LOAD'$ shows the updated load profile without considering updated memory load exceeding the limit in the first place; $LOAD''$ shows the updated load that guarantees its peak never exceeds the limit of 60 MB, where a certain operation has to be delayed. 

\section{Implementation}
\label{implementation}

In this section, we introduce the unified program abstraction for implementing SmartPool and AutoSwap, which could be adopted into an existing deep learning framework with ease. The core abstraction is the \emph{Device}  class, which is shown below:

\begin{verbatim}
class Device {
  Block* Malloc(size_t);
  void Free(Block*);
  void Exec(Function, vector<Block*>, 
    vector<Block*>);
  
  SmartPool PoolOpt();
  map<int, Event> SwapOpt();
  
  SmartPool pool;
  map<int, Event> schedule;
}
\end{verbatim}

The \emph{Device} class represents a computing device, i.e., a GPU. For every variable that uses GPU memory, its memory (represented by \emph{Block}) is managed via \emph{Malloc} and \emph{Free}. At the beginning of a training process when the memory pool is not created yet, these two functions are implemented by calling the native \emph{cudaMalloc} and \emph{cudaFree} respectively. Meanwhile, malloc/free requests are recorded into a list for detecting an iteration. Once two consecutive subsequences are detected to be repeating, the subsequence is fed into \emph{PoolOpt} for constructing the memory pool. This subsequence represents all operations involved in one training iteration. \emph{PoolOpt} extracts the lifetime of each variable and runs the SmartPool algorithm from Section \ref{allocation} to create a \emph{SmartPool} instance, i.e., \emph{pool}. \emph{pool} maps from each operation index (with a malloc request) to a GPU memory address for memory allocation. Upon creation of \emph{pool}, \emph{Malloc} and \emph{Free} will execute according to the map (implemented as a hash table).

In the meantime, all arithmetic operations running on the GPU device are executed by function \emph{Exec}, including cuDNN operations\footnote{cudnn operations are provided by Nvidia's cuDNN library (\url{https://developer.nvidia.com/cudnn})} and user-defined CUDA\footnote{https://developer.nvidia.com/cuda-toolkit} kernel operations. Two lists of variables are passed to \emph{Exec}, one for the variables to be read and the other for the variables to be written. Consequently, these read/write requests are recorded along with the malloc/free requests in a list which undergoes repeatability test as well for applying the AutoSwap approach. Then \emph{SwapOpt} is called to optimize the swapping schedule using algorithms from Section~\ref{auto-swap}. Note that before swapping schedule is made, the system swaps out variables as many as necessary when it is about to exceed the limit, and swaps them in when a variable that is not on the GPU memory is accessed again. No prefetching is applied. Hence, it would be slow in the early iterations. The swap-out and swap-in events are executed with 2 separate cudaStreams along with the default cudaStreams for computing. Swapping and synchronization are based on the swapping scheduling algorithm illustrated in Section \ref{scheduling}.

To combine SmartPool and AutoSwap, we create AutoSwap at first and then create SmartPool. The order cannot be exchanged because SmartPool depends on the sequences of \emph{Malloc} and \emph{Free}, and \emph{AutoSwap} calls additional \emph{Malloc} and \emph{Free} due to swap-out and swap-in.

\section{Evaluation}
\label{evaluation}

In this section, we evaluate the performance of SmartPool, AutoSwap, and the combined approach, i.e. the mode when the orthogonal SmartPool and AutoSwap approaches work concurrently. We conduct the experiments on a workstation equipped with CPU Intel Xeon E5-1650 v4 and GPU GeForce GTX 1080 Ti with 11 GB DDR5 RAM. The CPU memory is 64 GB DDR4 RAM. The motherboard is ASUS X99-E WS, which gives 16 PCI-E for data communication between CPU memory and GPU memory. The environment is Ubuntu 16.04 with CUDA 8.0 and cuDNN 5.1. We choose the fastest cuDNN algorithm (without workspace limit) for all the experiments. We use the image dataset CIFAR-10 \cite{cifar10}, and evaluate training performance using commonly used benchmark DNNs ResNet\footnote{ResNet: http://torch.ch/blog/2016/02/04/resnets.html} and VGG\footnote{VGG: http://torch.ch/blog/2015/07/30/cifar.html} of different depths.

Firstly, we evaluate the competitive ratio and the time complexity of SmartPool compared with the baselines CnMem Pool and cudaMalloc. Secondly, we evaluate the memory load reduction and the corresponding overhead of AutoSwap by using different PS and BO. Thirdly, we evaluate the scalability of the combined approach by comparing its memory footprint with CnMem Pool and cudaMalloc for different DNNs at various batch sizes. Finally, we compare the performance of our combined approach with 3 other memory reduction baselines.

\begin{table}[hbtp]
    \centering
    \caption{Competitive Ratio and Time Complexity of SmartPool, CnMem Pool and cudaMalloc}
    \includegraphics[width=1.0\linewidth]{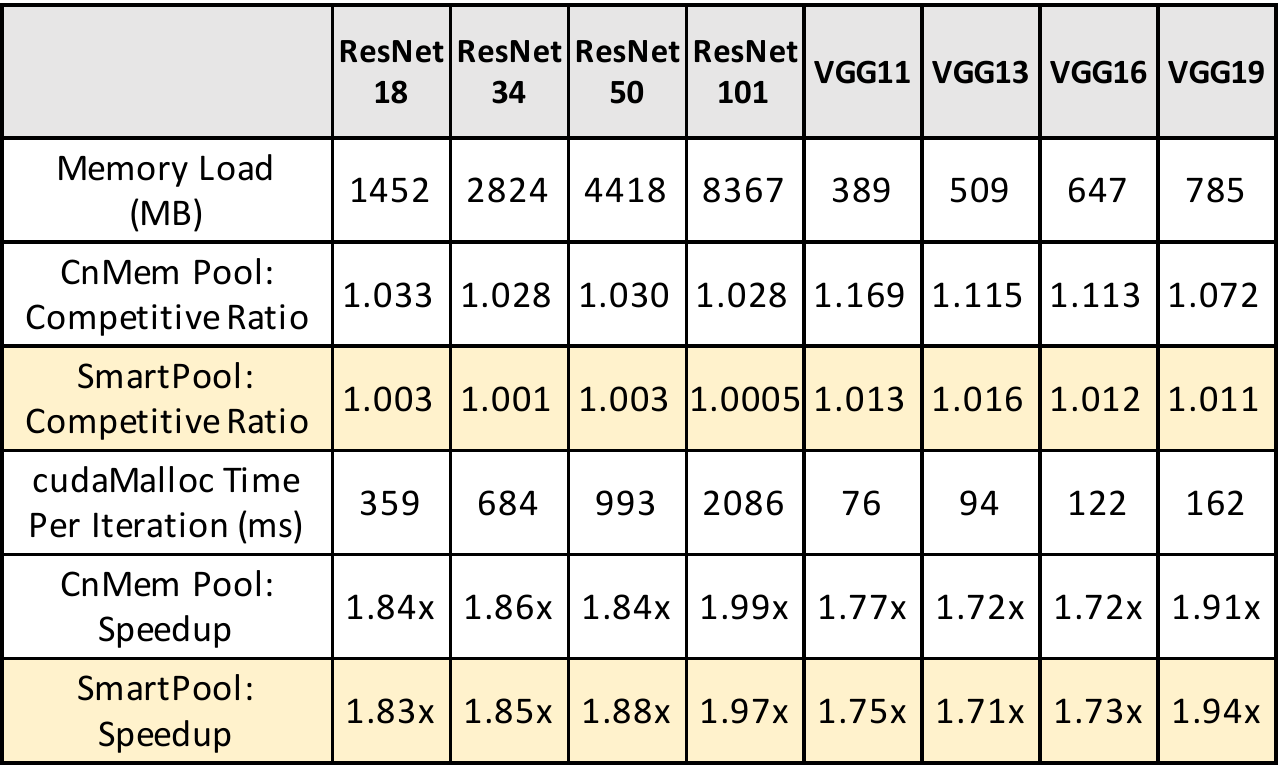}
    \label{tab:table_pool}
\end{table}

\subsection{SmartPool}
\label{eval-smartpool}

Our experiments show that first-fit and best-fit are comparable in terms of memory footprint. Thus, we directly compare the performance of SmartPool, Nvidia's native cudaMalloc, and its default memory pool CnMem Pool. We conduct the experiments on ResNet and VGG of different depths at a batch size of 100, with results shown in Table \ref{tab:table_pool}.

cudaMalloc consumes the exact amount of memory when allocating each and every variable. Hence, it gives optimal memory consumption which is equal to the peak memory load. However, frequently calling the cudaMalloc inevitably increases the time complexity of training and causes fragmentation due to various variable sizes. In this experiment, we take cudaMalloc as a baseline, and compare our SmartPool with CnMem Pool in terms of competitive ratio $\alpha$ and speedup. According to the definition of $\alpha$, the $\alpha$ of SmartPool and CnMem Pool here equals to their respective memory footprints over the footprint of cudaMalloc, the closer to 1 the better; speedup refers to the relative training speed compared to cudaMalloc, with higher value correspond to a shorter time spent per training iteration.

SmartPool achieves a very low $\alpha$ compared to CnMem Pool for all the network models in the experiment. Comparing $\alpha$ for VGG11, SmartPool reduces up to 13.3\% of memory footprint compared with CnMem Pool. SmartPool under the worst condition (VGG13) only introduces 1.6 \% fragmentation.

Exploitation of the iterative nature in training is the key to achieve near-optimal $\alpha$. Firstly, it reduces the online DSA to an offline one, allowing smarter memory allocation based on the lifetime and sizes of variables. Secondly, as discussed in Section \ref{allocation}, SmartPool enables superior memory sharing such that memory of a large variable can be shared with several smaller ones, and vice versa. It provides efficient memory sharing and reduces the memory footprint. Thirdly, our algorithm colors the variables with decreasing size, and hence the largest variables are allocated at the bottom of the memory pool, which also contributes to low $\alpha$ even for networks with large variables like VGG.

Concerning the time complexity, both CnMem Pool and SmartPool are much faster than cudaMalloc with up to two times speedup. In SmartPool, Malloc is implemented by looking up a table (C++ std::map) of n variables, with time complexity is $O(log\ n)$; in CnMem Malloc is done by searching over a linked list of m holes with time complexity $O(m)$, where $m<n$.

\subsection{AutoSwap}
In this subsection, we evaluate the memory load reduced by AutoSwap. Note that the evaluation does not concern the memory footprint. We start by comparing the communication overhead using different priority scores under various memory load limits for VGG16. We then analyze the optimality of the method. Finally, we present the results for other DNNs.

\subsubsection{Minimum Memory Load, $load_{min}$}

\begin{figure}
  \includegraphics[scale=0.55]{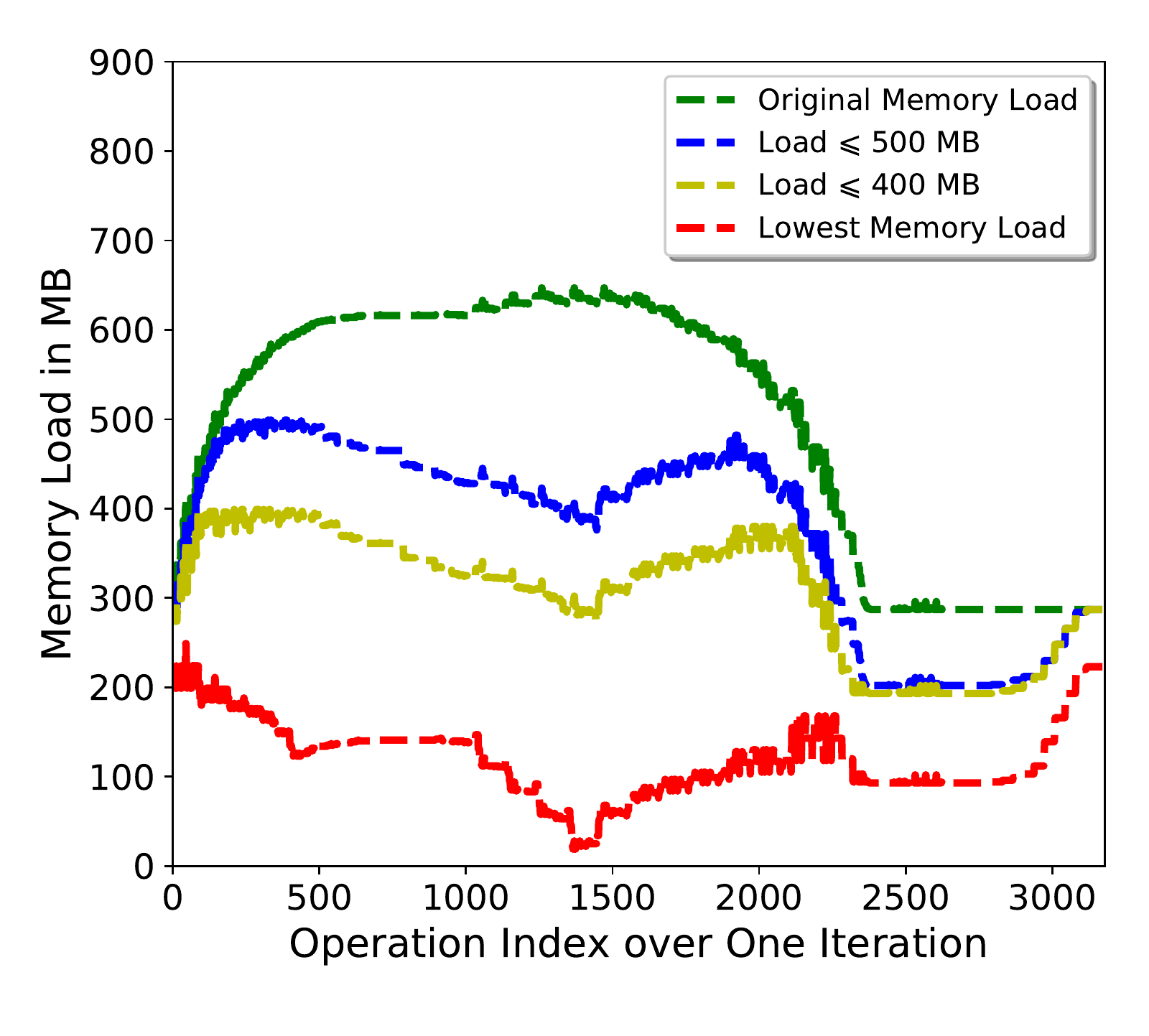}
  \caption{Memory load of VGG16 under different conditions.}
  \label{fig:lowest_load}
\end{figure}

Fig. \ref{fig:lowest_load} shows the memory load profile of VGG16 under different conditions: (i) original load without swap, (ii) updated load with swap at load limit of 500 MB and (iii) 400 MB respectively, and (iv) lowest load profile that is achieved by swapping out and in all candidate variables synchronously. Note that the peak at the 4th curve is located at the start of the iteration (or at the end of the iteration in other cases), the location of which is different from the original peak. We define this peak value as the minimum memory load, $load_{min}$. AutoSwap can swap memory load efficiently when the user-defined limit is above $load_{min}$, as it only needs to swap once for each portion of the memory load. If in the extreme scenario where the user-defined limit is below $load_{min}$, we can still achieve it by swapping variables not only from the candidate variables. In other words, a certain part of the memory needs to be swapped more than once and hence it is less cost-effective. As different DNNs vary significantly in $load_{min}$, we only evaluate the communication overhead for memory load limit above the $load_{min}$ for each DNN.

\begin{figure}
  \includegraphics[scale=0.57]{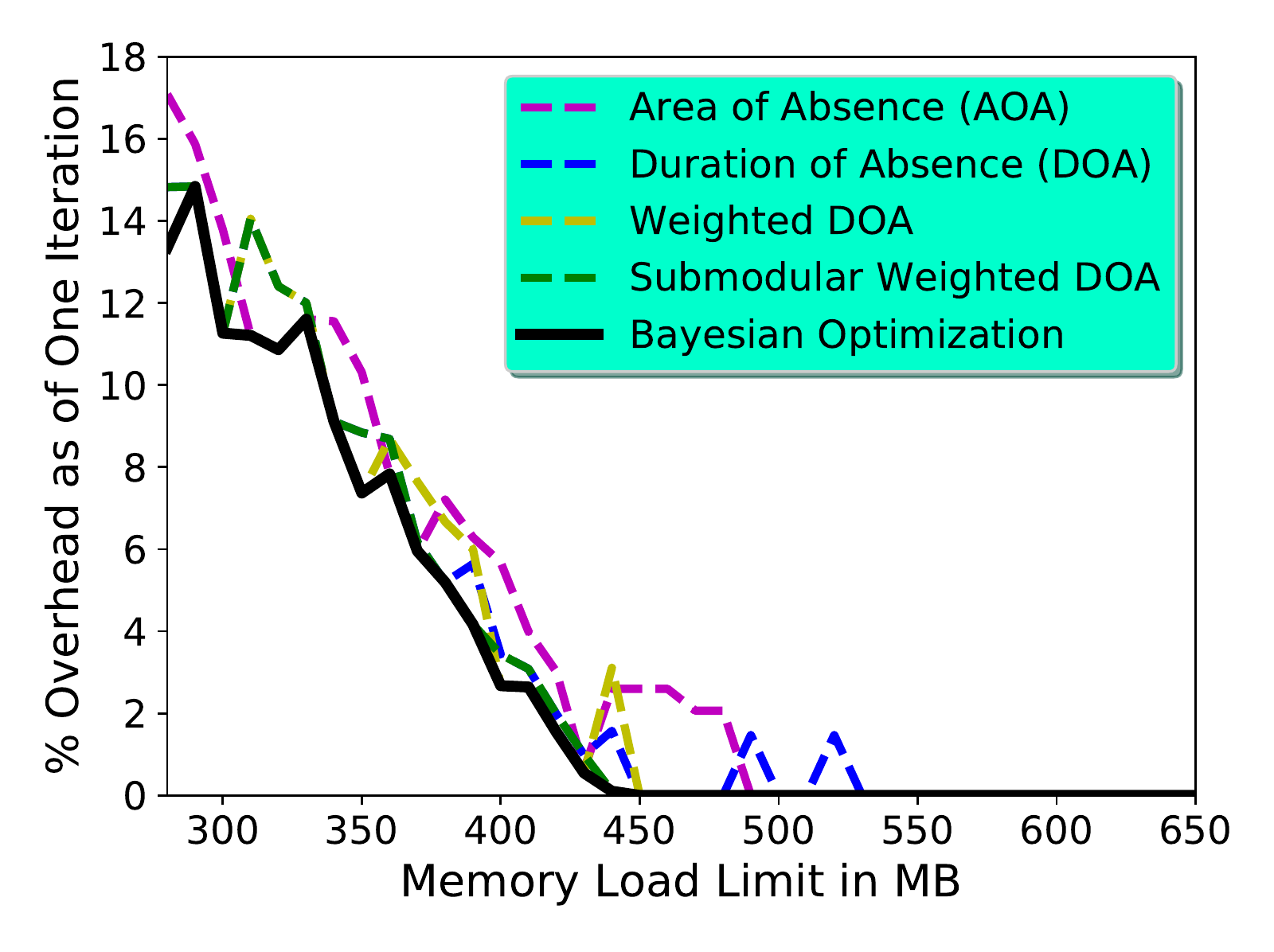}
  \caption{Overhead of different priority scores and Bayesian Optimization for VGG16.}
  \label{fig:overhead}
\end{figure}

\subsubsection{Comparing the Priority Scores}
We compare the communication overhead for the 4 PS and BO under the network VGG16 at a batch size of 100. Fig. \ref{fig:overhead} shows the comparison results. The results are across a wide range of memory load limits from 647 MB to 280 MB (slightly above the $load_{min}$). Comparing the 4 PS, we observe that WDOA and SWDOA exhibit lower overhead under most of the memory load limits, while under certain limits AOA or DOA performs better. The performances of these greedy methods differ by less than 3\% overhead as of one training iteration. However, it does not show a clear trend whether at a certain regime one method is better than the rest. All the methods can maintain a reasonable overhead of less than 20 \% throughout the experiment. In contrast, the use of BO safeguards the overhead to be no larger than the minimum of the 4 PS. Since there are only 4 hyperparameters to optimize, the time complexity is rather reasonable and it is able to converge within 30 to 40 evaluations. As a result, the system can achieve zero overhead with the memory load reduced to 447 MB, which is about 30\% of memory load reduction without increasing the training time. It can be noted from Fig. \ref{fig:overhead} that the overhead does not always monotonically increase with lower memory load. This is in fact due to the granularity of candidate variables. 

\subsubsection{Optimality Analysis}
We now zoom into the conditions with overheads, to analyze if some portion of the overhead is avoidable by any chance. We observe that the overhead in most cases occurs during forward-propagation, where memory load increases significantly faster than backward-propagation, as shown in Fig. \ref{fig:lowest_load}. If there is any time when the swap-out cudaStream is idle during forward-propagation, we explore if any variables that were not selected could be swapped during this time and hence indirectly avoid the overhead contributed by other variables. 

For example, the memory load of VGG16, due to more variables being created in forward-propagation, takes 24 ms to ramp up to 95\% of the peak memory load from the start of an iteration. Swapping out 300 MB to make the memory load below 350 MB takes around 28.9 ms, which incurs 7.3 ms overhead by the BO method, out of which only less than 1.7 ms is avoidable overhead. In another word, our algorithm manages to hide more than 94\% of the communication that is avoidable in the background. 

We conduct this analysis of various models under different memory load limits. We observe that in most cases there is no avoidable overhead or only a negligible amount of overhead is avoidable. It indicates that our greedy algorithm is indeed near optimal. 

Moreover, it is worth pointing out that there are a number of variables whose lifetime and two consecutive accesses are cross one or more iterations, such as weights that are kept till the end of the process. They are more favorable for swapping which can be swapped out at one iteration after the last access and swapped back at the next iteration before the first access. This provides wider time window such that the communication is fully hidden under computation. Our PS and BO successfully capture them which significantly maximize the PCI-e bandwidth and hence makes the overhead ideally low.

\begin{table}[hbtp]
    \centering
    \caption{Maximum Memory Load Reduction without Overhead}
    \includegraphics[width=0.9\linewidth]{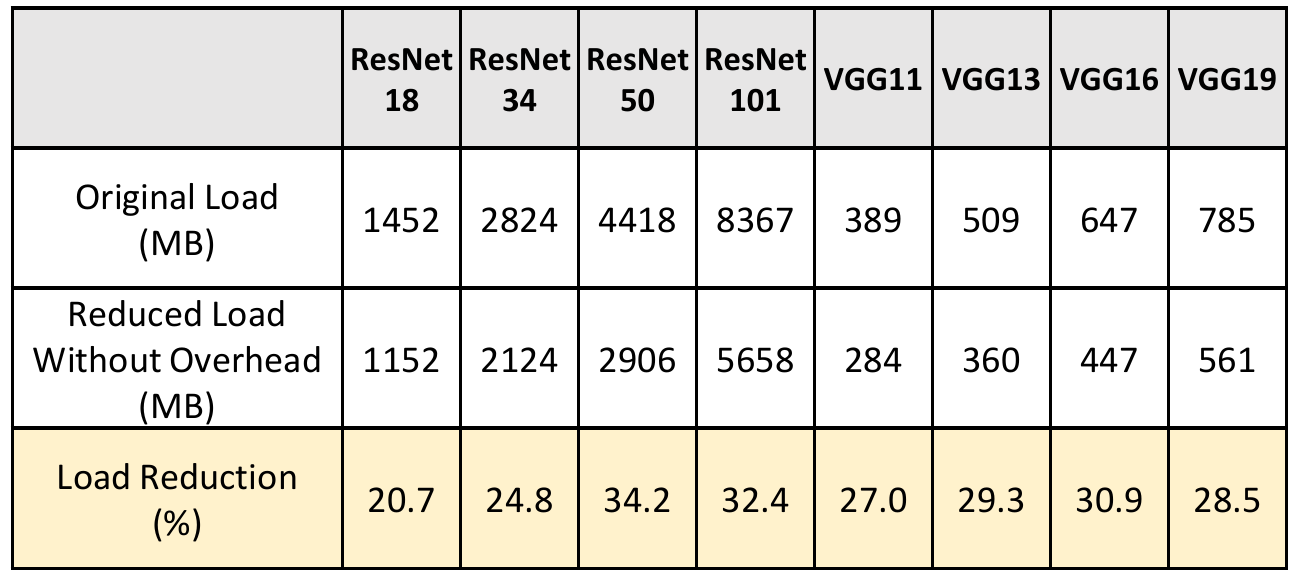}
    \label{tab:reduction}
\end{table}

\subsubsection{Overhead of Various Models}
In this section, we evaluate other models by the BO method. Table \ref{tab:reduction} shows the maximum memory load reduction without overhead under different network depths of VGG and ResNet using the same batch size of 100. For VGG network structure, it reduces up to 30.9 \% memory footprint without overhead, from 647 MB to 447 MB in VGG16. For ResNet network structure, it reduces up to 34.2 \% memory footprint without overhead, from 4418 MB to 2906 MB in ResNet50. We also evaluate the overhead under different memory limits for each model, which are presented in Section \ref{combined-approach} (Fig. \ref{fig:memory_vs_overhead}), where the $x$-axis is the footprint rather than memory load. The swapping performance shows considerable scalability for both different memory limits and different depths of networks.

\subsection{Combined Approach}
\label{combined-approach}
\subsubsection{Comparisons with cudaMalloc and CnMem Pool}

\begin{figure}
  \includegraphics[scale=0.57]{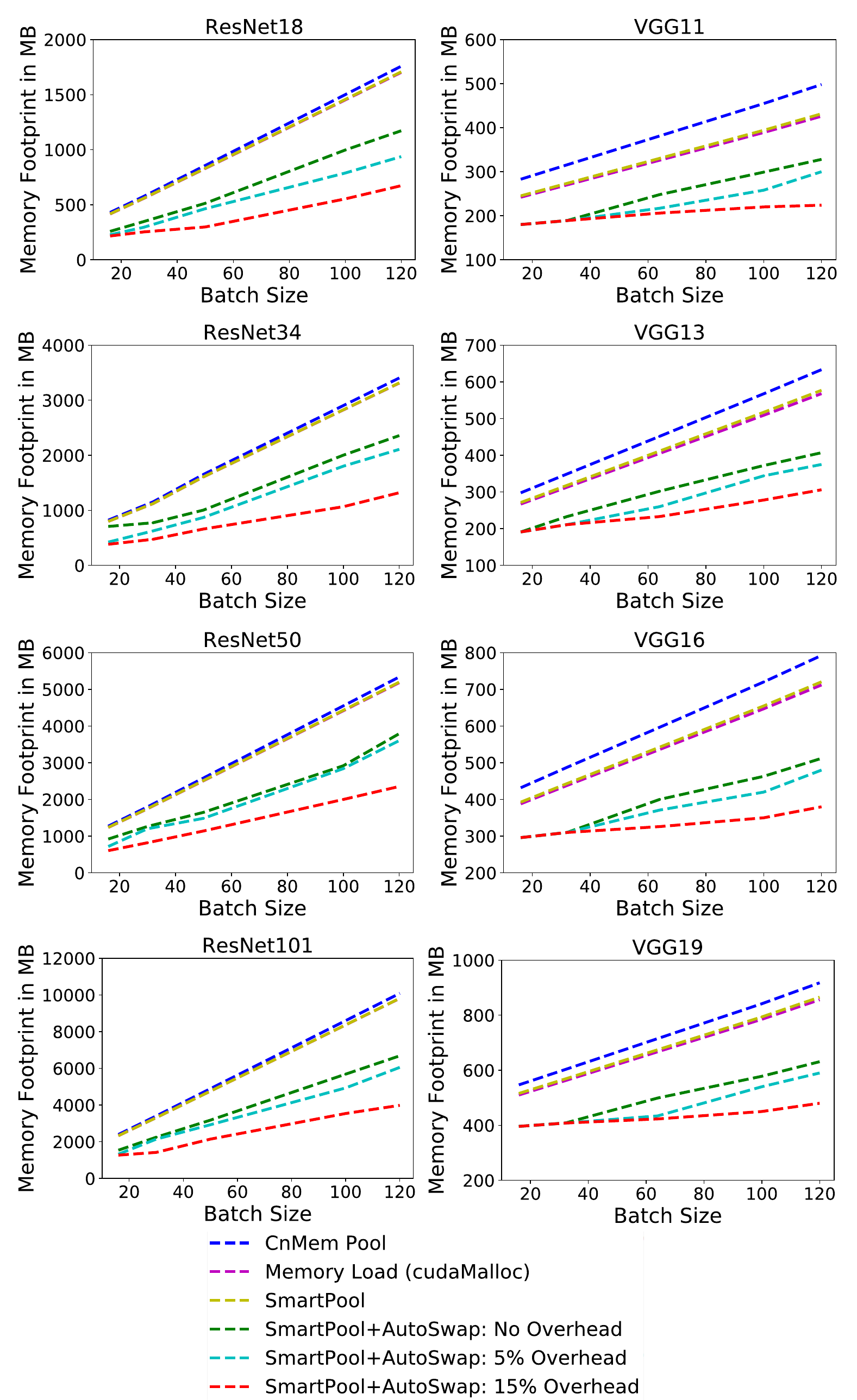}
  \caption{Memory footprint of different memory management schemes at various batch sizes.}
  \label{fig:footprint_vertical}
\end{figure}

Now, we consider both the allocation of variables and the memory load reduction. With the SmartPool and AutoSwap applied simultaneously, we perform experiments to evaluate the memory footprint reduction under different network depths and different batch sizes. Fig. \ref{fig:footprint_vertical} compares the memory footprint of cudaMalloc, CnMem Pool, SmartPool, and the combined approach SmartPool+AutoSwap. We take the footprint of CnMem Pool as the baseline. At each batch size for each model, we can see the clear gap between the footprint of CnMem Pool and the combined approach with different values of overheads. It shows that we can reduce up to 1/3 of the footprint without increasing the training time, and reduce no less than 60\% of the footprint with smaller than 15\% overhead. 
The performance is scalable when we increase the depth of the network and use a larger batch size. The percentage of memory reduction of the combined approach with 15\% overhead does not decrease when we increase model depth. The curve for each network is linear with a smaller slope than the baseline, which shows that it works well in deeper networks and even more promising with larger batch sizes.
\subsubsection{Comparisons with Other Baselines}
Finally, we compare our combined approach with three other memory reduction methods as baselines. They are MXNet-memonger based on MXNet \cite{chen2016training}, GeePS based on Caffe \cite{cui2016geeps}, and a recent deep learning framework SuperNeurons \cite{wang2018superneurons}. MXNet-memonger uses memory sharing, in-place operation, and trading computation for memory. The optimization is performed automatically. However, the performance of memory reduction depends on the choice of searching nodes for recomputing. SuperNeurons provides options to recompute and/or swap, where the swapping is restricted to convolution layers. It uses memory pool which is only for efficient allocation and deallocation of variables and does not optimize the memory allocation inside the memory pool. GeePS is a distributed parameter server system which solely uses memory sharing. This method allows us to choose different GPU memory limitation and the decision of which layer or which tensor to swap is made by the end user.

Since different frameworks vary significantly in the absolute value of computation time and memory consumption, we present the results with the percentage of overhead incurred as of one iteration and the percentage of memory footprint reduction with memory optimization option. To compute the percentage of the overhead, we measure the average time per iteration with and without memory optimizing option for each baseline. Memory consumption can be measured from nvidia-smi at the microsecond level, except for MXNet-memonger, which takes up more GPU memory at the beginning for searching the optimal convolution algorithm. Hence, its actual memory consumption is measured from the steady-state of the training phase.

We managed to cover the widest possible range of data points for the baselines and our combined approach. The combined approach is able to give a wide range of memory consumption from baseline footprint to $load_{min}$. For MXNet-memonger, we change the position of searching nodes at different layers and unit blocks, and obtain 3 data points for each network. For SuperNeurons, we can have options to recompute and/or swap, which provides 3 different combinations of compilation and hence 3 data points. For GeePS, we run the training model just on one machine and one GPU instead of a distributed environment for fair and accurate comparison. It gives a wide range of memory consumption; the baseline time and memory consumption are obtained using original Caffe. The batch size is fixed at 100 for CIFAR-10 dataset using ResNet and VGG of various depths. For those networks that are not able to be fit in our GPU (11 Giga Bytes) under certain baselines, no data point is available.

\begin{figure}
  \includegraphics[scale=0.57]{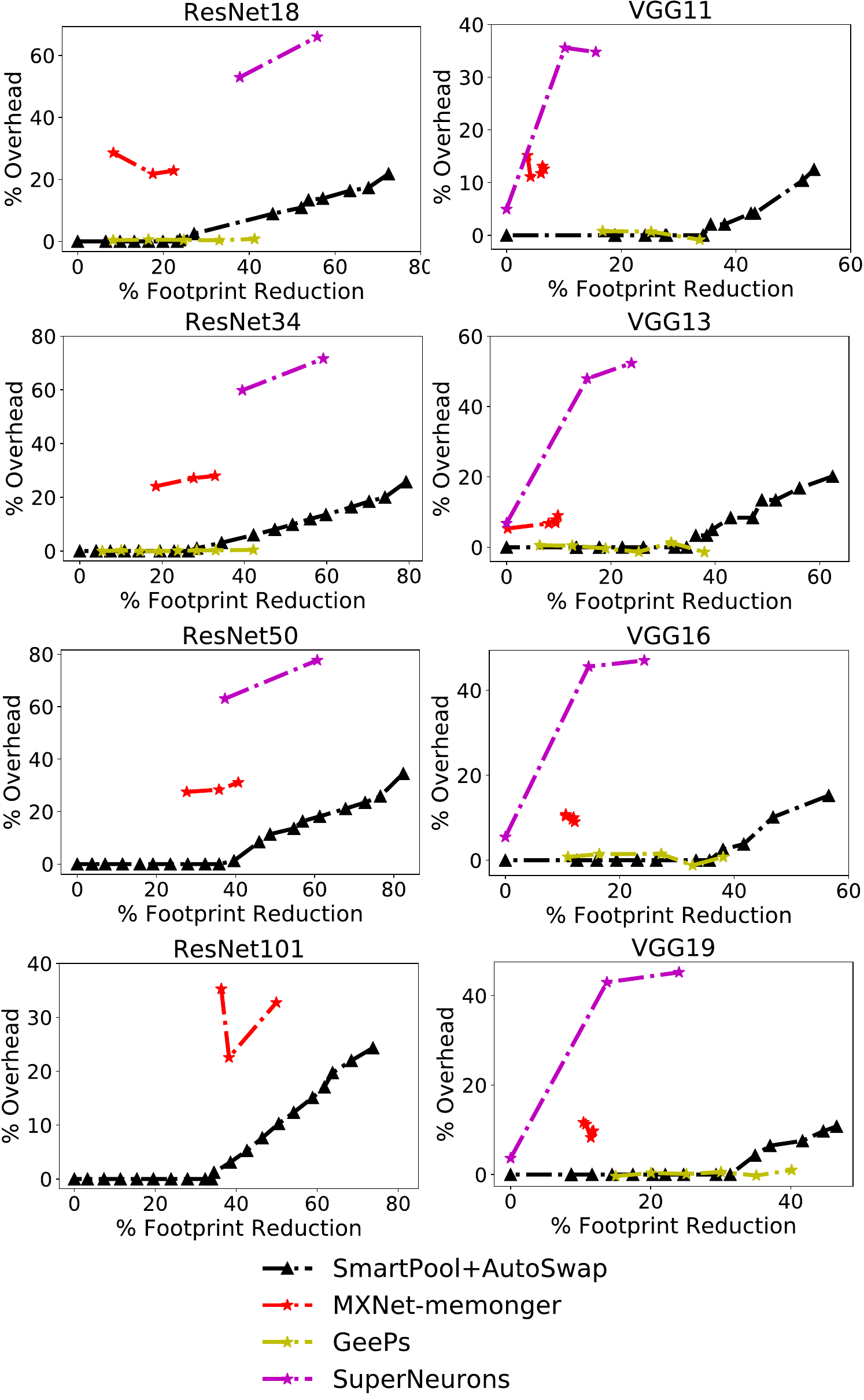}
  \caption{Memory footprint reduction and overhead for different depths of VGG and ResNet models.}
 \label{fig:memory_vs_overhead}
\end{figure}

The percentage of overhead versus the percentage of footprint reduction for all the four methods are shown in Fig. \ref{fig:memory_vs_overhead}. MXNet-memonger shows significant memory reduction for VGG models, reducing 40\% of the memory consumption with only less than 30\% of overhead. However, it has less effect on ResNet models. SuperNeurons is able to reduce more memory consumption than MXNet-memonger by recomputation and swapping. However, its overhead is constantly higher than that of the MXNet-memonger. GeePS exhibits zero or nearly zero overhead within a wide range of footprint. In overall, the overhead of GeePS is comparable with that of SmartPool+AutoSwap. Under some conditions, its overhead is smaller than our combined approach. However, the memory consumption of GeePS cannot be reduced further to what can be achieved by SmartPool+AutoSwap, as cudaMalloc would be out of memory in constructing the network when the memory limit is below a certain value for certain models. Moreover, users of GeePS have to manually decide which layers and tensors to swap, and hence the actual performance depends on the skills and knowledge of the end user.

Compared to other methods, our combined approach has the following advantages: Firstly, it provides adjustability on the percentage of footprint reduction within a wide range. It meets the requirement while not swapping memory excessively, and hence avoiding redundant data communication. Secondly, our approach is transparent to users and starts working automatically at the early iterations. It does not require knowledge of the user on the memory consumption of the network, or deciding which layer to swap, etc. Thirdly, it gives low overhead with considerable scalability as we vary the DNN type and depth, batch size, and the percentage of footprint reduction.  

\section{Conclusion and Future Work}
\label{conclusion and future work}
In this paper, we exploit the iterative nature in training DNNs and propose two orthogonal approaches, SmartPool and AutoSwap, to reduce the GPU memory consumption efficiently and effectively. They are transparent to the end users and do not require human intervention. Experiments show that SmartPool effectively optimizes the allocation of variables in the memory pool; AutoSwap efficiently reduces the memory load by swapping out the currently not-in-use variables to CPU memory with ideally low communication overhead; the combined approach further reduces the memory footprint. In addition, it scales well for different network architectures as we vary the network depth, the batch size, and the memory load limit.

In the future work, we will extend our solutions for the applications whose iterations have slight variations, explore memory reduction in a distributed environment~\cite{jiang2017heterogeneity}, and adapt our approaches into other memory-hungry applications with iterative nature, such as large-scale K-Means running on GPU.

\bibliographystyle{IEEEtran}
\input{pub.bbl}

\end{document}